\documentstyle[epsfig]{elsart}

\begin{document}
\begin{frontmatter}

\title{Ballistic reflection at a side-gate in a superconductor-semiconductor-superconductor
 structure}
\author{G.~Bastian and H.~Takayanagi}
\address{NTT Basic Research Labs, 3-1 Morinosato Wakamiya, Atsugi,
Kanagawa 243-0198, Japan}

\begin{abstract}
We have fabricated a sub-micron-sized structure consisting of an
InAs-based 2DEG, two narrow Nb leads and a gate, where the
indirect ballistic transport between the non-oppositely
superconducting contacts can be controlled by the voltage applied
to the gate. This new kind of tuneable junction can be used for
applications and allows several fundamental questions related to
the transport mechanism to be studied. First results of
experiments carried out in this respect are presented.
\end{abstract}
\end{frontmatter}

pacs[74.40+k,74.50.+r]

\section{Introduction}

Controllable Josephson weak links using semiconducting material
as a barrier were used to investigate fundamental transport
mechanisms as well as possible applications. By influencing the
carriers inside the semiconductor, Josephson field effect
transistors \cite{Clark80}, non-equilibrium junctions
\cite{Testardi71,Morpurgo98,Baaselmans99} or optically modulated
weak links \cite{Bastian99,Schapers99} were demonstrated. In this
article we present a novel device, where the transport of
ballistic carriers can be controlled by a tunable side-gate.

\begin{figure}[htb]
\begin{center}
\epsfig{file=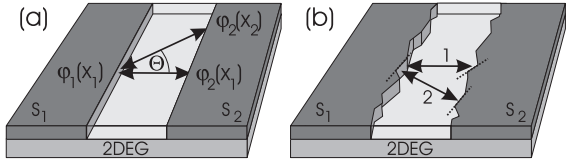,width=7.5cm}
\end{center}

\caption{(a) Andreev bound states can be formed between different
positions and phase differences along the width of the interface.
(b) In the case of rough interfaces, the shortest connection
(path 1) can have a lower probability than longer connections
(path 2) due to a more favorable angle of incidence.}
\label{nonlocal}
\end{figure}

Only recently, more realistic descriptions have been provided of
devices where superconductors are attached to a 2DEG and all
dimensions are comparable to the phase-breaking length
\cite{Mortensen99}: as depicted in Fig.\ref{nonlocal}a, taking
into account a realistic geometry with two dimensions in which
carriers can transfer from one superconducting contact to the
other, a modified subgap structure \cite{Mortensen99}, different
density of states \cite{Chaudhuri95} and a different magnetic
field dependence are theoretically predicted consequences of these
{\it non-local modes}.

In Josephson junctions, the existence of these modes is related to
the angular dependence of the Andreev reflection probability and
can play a crucial role: whenever the roughness of the interface
between the superconductor and the barrier material is larger
than the Fermi wavelength of the quasiparticles inside the normal
conducting region, the transport is in fact neither perpendicular
nor restricted to one dimension any more, as shown in
Fig.\ref{nonlocal}b.

The angle dependent reflection probability at perfect interfaces
between two materials depends on the ratios of their carrier
densities and on their effective masses or, correspondingly, on
their Fermi velocities $v_F$ and wave vectors $k_F$. At a
non-idealized superconductor-semiconductor interface the total
barrier strength can be described by an effective angle dependent
parameter $Z_{eff}(\Theta)$ \cite {Mortensen99}, which allows the
iv-characteristic of a Josephson junction to be calculated more
accurately using a generalized OBTK model. However, the underlying
angular dependence has neither been verified experimentally nor
used for application.

\section{Samples}

\begin{figure}[htb]
\begin{center} \epsfig{file=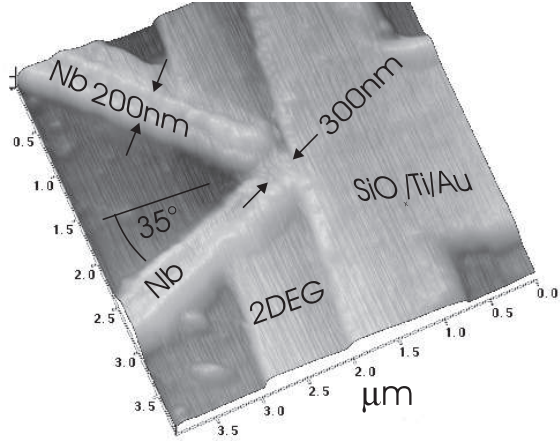,width=7.5cm} \end{center}

\caption{The effective geometrical electrode separation under
perpendicular injection ($\Theta=0$) is about $2\times300~$nm. In
order to resolve an angular dependence of Andreev reflection, the
superconducting leads have a rather rectangular shape and a width
of 200~nm only.} \label{AFM}

\end{figure}

In structures with gates previously studied, the carrier density
in the semiconducting barrier was modulated homogeneously and
locally \cite{Takayanagi99}, or small channels were defined
through which the superconductors are coupled
\cite{Takayanagi95}. In contrast to this, in the case of our
devices a potential barrier can be introduced at a certain
position within the 2DEG depending on the voltage $V_g$ applied
to the gate. At this position ballistic reflection of the
carriers occurs \cite{Comment03}. Quasiparticles which are
injected into the 2DEG from one superconductor can reach the
other superconductor only indirectly by undergoing normal
reflection at this induced potential barrier. An AFM picture of a
fabricated structure with which this effect can be demonstrated
is shown in Fig.\ref{AFM}. Local shifting the potential barrier by
a change of $V_g$ defines both, the angle at which the other
superconductor can be reached and the effective length of the
path.

The structure has been fabricated by dry and selective wet etching
of an InAs/AlSb/GaSb heterostructure. The Nb leads were
evaporated on top of the InAs layer after weak Ar plasma
cleaning. To avoid gate leakage, an evaporated layer of Ti/Au was
electrically isolated by a SiO$_x$ layer. Several sets of samples
with different angles and dimensions were prepared using 2DEGs
with a carrier concentration of $n_e=9.9\times10^{11}~cm^{-2}$
and mobility of $\mu=191000~cm^2/Vs$. This results in mean free
paths of 3.15~$\mu m$, which is much larger than the device's
dimensions so that the transport is far in the ballistic regime.
The alignment of side-gate, Nb-leads and etching masks with
respect to each other turned out to be the most delicate part.

\section{Results}

\begin{figure}[htb]
\begin{center}
\epsfig{file=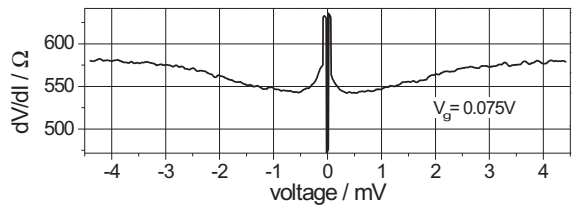,width=7.5cm}
\end{center}
\caption{Typical differential resistance at a temperature of
45~mK.} \label{dVdI}
\end{figure}

We measured the iv-characteristics and the differential resistance
of the device by a standard four-point technique between the two
Nb leads and determined its dependence on $V_g$. As can be deduced
from the $dV/dI(V)$ curve in Fig.\ref{dVdI}, where a sharp dip at
zero bias voltage and an excess current can be seen, a
superconductive coupling is formed \cite{Comment01}.

\begin{figure}[htb]
\begin{center}
\epsfig{file=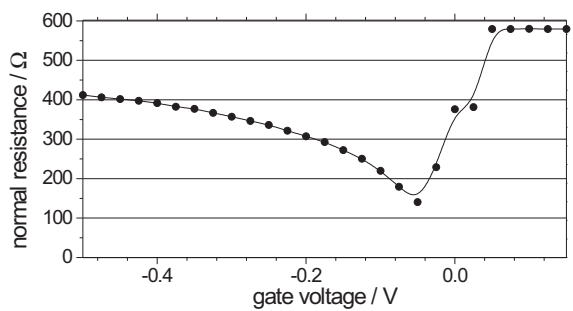,width=7.5cm}
\end{center}
 \caption{The normal resistance
shows a minimum value for small negative gate voltages and can be
altered by about 500\%.} \label{gate-dependence}
\end{figure}

In Fig.\ref{gate-dependence} the normal resistance of the device
is plotted against $V_g$. Above a certain positive voltage the
resistance reaches a constant maximum value. This corresponds to
the situation, where there is no induced potential barrier and
ballistic electrons are not reflected. The indirect transport from
one Nb-lead to the other is therefore very improbable
\cite{Comment02}. The occurrence of this offset in $V_g$ has been
reported several times \cite{Kristensen98} and is about
$V_g\approx+0.1~V$ in our case. As $V_g$ is reduced the normal
resistance decreases and reaches its minimum value at about
$V_g\approx-0.050~$V. The indirect transport via reflection at an
induced potential barrier decreases the resistance and emerges an
optimum when the corresponding path has the highest Andreev
reflection probability. A further reduction of $V_g$ slowly
increases the normal resistance again, as the angle of incidence
of injected quasiparticles becomes less favorable.

Thus the side-gate can be used to adjust the focus of the
transport of ballistic quasiparticles, which enables a large
change of the normal resistance. Best coupling could be achieved
at an effective gate voltage of -0.15~V and a resulting
theoretical maximum potential barrier height of 150~meV
\cite{Glazman91}, which is reasonable compared with the Fermi
energy of the 2DEG of $E_F\approx70~$meV. A quantitative model for
the effect observed must take the precise profile of the induced
potential, its rise and shift depending on $V_g$, and the
resulting possible transport modes inside the device into
consideration \cite{Mortensen00}.

\section{Summary}

In summary, we fabricated a new type of tunable ballistic
Josephson junction with the unique possibility of changing the
effective electrode separation. In addition, the angular
dependence of Andreev reflection and the formation of {\it
non-local modes} can be made use of and studied with this device.
A modulation of the normal resistance by up to 500\% could be
achieved.

\ack We should like to thank J.~Schmitz, M.~Walther, J.~Wagner
(IAF Freiburg) for the growth of the heterostructures and
N.A.~Mortensen for fruitful discussions. This work was supported
by NEDO.


\begin{thebibliography}{99}
\bibitem{Comment01}The exact value of the critical current could not be measured due
to rounding caused by different sources of noise.
\end{thebibliography}
\end{document}